# Phenotypic variability in Italian rice germplasm

Gabriele Mongiano[*,a], Patrizia Titone[a], Simone Pagnoncelli[a], Davide Sacco[a], Luigi Tamborini[a], Roberto Pilu[b], Simone Bregaglio[c]

[a]CREA - Council for Agricultural Research and Economics, Research Centre for Plant Protection and Certification, SS 11 km 2.5, 13100, Vercelli, Italy

[b]University of Milan, Department of Agricultural and Environmental Sciences - Production, Landscape, Agroenergy, Via Celoria 2, 20133, Milan, Italy.

[c]CREA - Council for Agricultural Research and Economics, Research Centre for Agriculture and Environment, via di Corticella 133, 40128, Bologna, Italy.

*Corresponding author:

**Abstract**

Plant breeding is one of the key strategies for further enhancement of crop yields; however, effective breeding strategies require phenotypic characterisation of the available germplasm. This study sought to characterise the phenotypic expression of fourteen crop traits related to phenology, plant architecture, and yield in a panel of 40 cultivars selected to represent the phenotypic variability present in Italian rice germplasm during a two-season field experiment. The observed range of phenotypic variation was high for many traits (coefficients of variation ranging from 5.9% to 45.4%) including yield (mean: 6.47 t ha$^{-1}$; CV: 15.4%; min: 2.19 t ha$^{-1}$; max: 8.95 t ha$^{-1}$), and multiple strong associations emerged in all the analysed traits. Cluster analysis extracted three groups of genotypes characterised by alternative yield strategies, i.e. "high-tillering", "early maturing", and "increased source-sink". Our findings highlight that rice yield decrease when one strategy is overemphasised. In contrast, the highest-yielding genotypes have a balanced ratio between sink and source organs, along with a proportional duration of vegetative and reproductive phases. This study depicts the phenotypic variability in Italian rice cultivars and proposes a novel classification based on yield-related traits which could be of use in multiple rice breeding applications.

# 1 Introduction

After the tremendous increases reached during the green revolution, the yields of the primary cereal crops - rice, wheat and maize - are now stagnating (Brisson et al., 2010; Lobell et al., 2011; Van Wart et al., 2013). New strategies are needed to assist breeding programs to further enhance crop productivity, especially in light of the detrimental impacts of climatic changes on food security and on the environmental sustainability of cropping systems (Bocchiola, 2015; Diffenbaugh and Giorgi, 2012). Among the available technologies, plant breeding is regarded as one of the critical strategies to sustain agricultural production and implement effective adaptation strategies (Zaidi et al., 2019). Plant breeding is indeed an expensive, time consuming and labour-intensive activity (Acquaah, 2012), which is limited by constraints like the narrowed genetic diversity and the poor adaptability of available genotypes across environments. Scientists are thus developing new techniques to speed up and lessen the economic burden of standard breeding programs like rapid generation advance (Lenaerts et al., 2019), speed breeding (Watson, 2018), genetic editing (Song et al., 2016), ideotype breeding (Martre et al., 2015), and crop simulation modelling (Eeuwijk et al., 2019; Hammer et al., 2016). All these techniques rely on a thorough knowledge of the accessible genetic variability and of the complex relationships between the crop traits of interest. Notably, the adoption of crop simulation modelling to support plant breeding programs (Boote et al., 2001; Chenu et al., 2011) should account for the known physiological limits and the correlations and compensatory effects among traits (Picheny et al., 2017). However, the costs associated with the characterisation of available germplasm through field experiments often impede the collection of exhaustive datasets needed for such activities (Nwachukwu et al., 2016).

These considerations also apply to rice agriculture in Italy, where there is a long history of cultivation and vast biodiversity of genotypes (Mongiano et al., 2018). Since the 1990s, the rice sale price has fallen considerably due to the competition exerted by developing countries (Food and Agriculture Organization, 2012). Moreover, the EU Common Agricultural Policy endorsed a progressive reduction of the subsidies granted to rice growers aiming at more equable income support to farmers regardless of the cultivated crops (European Parliament, 2010). Rice cultivation in Europe is also in the spotlight for ecological issues like groundwater pollution, high greenhouse gases emissions (mostly methane, W. Kim et al., 2018), and land degradation (Blengini and Busto, 2009), although it provides socio-economic benefits like water catchment (e.g. rice paddies used as floodplains) and the creation of habitats for waterbirds in lowland areas like, e.g., in the Italian Po valley (Fasola et al., 1996; Longoni, 2010). Apart from the ecological benefits, the environmental costs of rice cultivation and the use of prime agricultural land must be at least justified by its economic return, while actively fostering the reduction of the impacts of these agroecosystems (Sheehy et al., 2011).

The availability of phenotypic characterisations of the accessible germplasm is of paramount importance for the development of new improved rice varieties, which could bolster the rice systems towards greater economic and environmental sustainability. To date, assessments of the phenotypic variability found in the rice germplasm of Italy are lacking or involve a limited set of crop traits (Faivre-Rampant et al., 2010; Mongiano et al., 2018). To bridge this gap, we analysed fourteen key crop traits related to phenology, development, plant and grain biometrics, and biomass accumulation in a sample of 40 rice cultivars, selected to be representative of the whole phenotypic variability of the Italian varietal landscape. The aims of this study were a) to broaden the knowledge about the variability of yield-related traits, b) to investigate the between-traits relationships, and c) to highlight the patterns of similarities among Italian rice cultivars.

## 2  Materials and methods

### 2.1  Plant material

The 40 rice cultivars were chosen from a collection of 351 genotypes, which were characterised for seven crop traits in a previous study (Mongiano et al., 2018) – i.e., duration of vegetative and reproductive stages, culm and panicle length, caryopses length, width, and weight. The cultivar selection was performed via Kennard-Stone algorithm (Kennard and Stone, 1969) to obtain a sample maximising the variance of known traits and including genotypes at the tails of the distributions. We assumed that a high variation in these traits would be reflected in the variability of other crop traits, because of multiple internal correlations. The resulting sample included rice cultivars released during along the 20$^{th}$ century, with few cultivars established in early 1900 (Americano 1600, Originario, Fortuna, Senatore Novelli) and many modern varieties (Dante, Reperso, Megumi, Brezza, Carnise precoce, Meco), also including Clearfield® (*imidazolinone*-tolerant) genotypes (Terra CL, Leonidas CL).

### 2.2  Experimental conditions

A two-year (2016 and 2017) field trial was carried out in Northern Italy (45°19'16.8"N, 8°21'35.6"E, Vercelli), adopting the typical farmer management in the area (i.e., direct seeding and flooded conditions with drainings for chemical weeding and mineral fertilisation). The agronomic practices aimed at avoiding any yield loss due to plant pests and pathogens (Supplementary Table S1). We used a completely randomised design with two replications, with the experimental unit consisting of a 14.4 m$^2$ plot (six rows, interspaced 0.2 m and 9 m long). The trials were seed drilled on April 27$^{th}$, 2016 and April 24$^{th}$, 2017, with seed rates adjusted to reach 1500 viable seeds for each plot, according to the specific germinability of the seeds and thousand seed weight of the different cultivars.

The thermal conditions in the two growing seasons are presented in Supplementary Figure S1. Season 2016 was characterised by below-average temperatures in May and in the last decade of

June. In July and August temperatures were in line with the climatic norm, favouring a gradual recovery of the initial delay of vegetative phases. The month of September was warmer than in the last ten years, with an average increase in maximum temperatures of about 2 °C. Rainfall was among the lowest recorded in recent years, especially in April, August, and September, with precipitations about 20% lower than the average of the last decade. The year 2017 was also characterised by a sharp reduction in precipitations, which in the period from April to October were 60% less than in the previous decade. Monthly maximum temperatures were higher than in the last ten years, especially in June and October (2 °C higher), while monthly minimum temperatures did not vary significantly. The highest temperatures were reached in the first week of August, with peaks between 35 °C and 40 °C.

## 2.3   Data collection

Fourteen yield-related traits (Table 1) were selected from literature and measured in the field trial. The rationale for the choice of the traits was threefold: having been measured in previous phenotypic characterisation (Mongiano et al., 2018), the evidence of their direct or indirect link with yield (Fageria, 2007; Katsura et al., 2007; Samonte et al., 1998; 2001; Wu et al., 1998; Yang et al., 2018), and their representation in crop simulation models as output variables or crop parameters (Confalonieri et al., 2009; Jamieson et al., 1998; Li et al., 2017; Stöckle et al., 2003; Tang et al., 2009; Van Diepen et al., 1989).

Table 1 – List of the considered traits with the notation of abbreviations used, measure unit, and sample size.

| Code | Trait | Unit | Sample size | Notes |
|---|---|---|---|---|
| Phyl | Average phyllochron | °C day$^{-1}$ | 5 plants | Average value |
| Heading | Days from sowing to 50% heading | days | Plot | Date determined at 50% heading |
| GDDflo | Degree days from emergence to 50% heading | °C day$^{-1}$ | Plot | Calculated from the date of heading |
| Maturity | Days from 50% heading to maturity | days | Plot | Date determined at complete final toning of the hulls |
| GDDmat | Degree days from 50% heading to maturity | °C day$^{-1}$ | Plot | Calculated from the date of maturity |
| CulmLen | Culm length | cm | 20 shoots | Measured from crown root to the panicle node |
| ShootDM | Shoot biomass at full flowering | g shoot$^{-1}$ | 5 plants | Average biomass of a single shoot (i.e. culm, leaves, sheats, and complete unfilled panicle) considering tillers and main stem |
| Density | No. of panicles per m$^2$ | panicles m$^{-2}$ | 0.5 m$^2$ | Measured in a random area in the plot before harvest |
| FinLeafNum | Final leaf number | unitless | 5 plants | Counted on the main culm at flowering |
| LAI | Leaf area index at full flowering | m$^2$ m$^{-2}$ | 10 measures | Estimated using PocketLAI app (Confalonieri et al., 2014) |
| PaniLen | Panicle length (main axis) | cm | 20 plants | Measured from panicle node to the tip |
| PaniDM | Panicle dry mass | g panicle$^{-1}$ | 10 panicles | Dry weight of a complete panicle |
| PaniBranches | No. of panicle branches | branches panicle$^{-1}$ | 10 panicles | Number of secondary branches in the panicle |
| GrainWeight | Grain weight | mg | 800 seeds | Weight of a single spikelet |
| GrainNum | No. of grains per panicle | grain panicle$^{-1}$ | 5 panicles | Total number of spikelets per panicle |
| Sterility | Sterile spikelets | % | 5 panicles | Ratio of sterile spikelets over the number of spikelets per panicle |
| Yield | Grain yield dry mass | t ha$^{-1}$ | Plot | Average dry weight of grain yield, expressed in t ha$^{-1}$ |

The fourteen crop traits were measured following the standard protocols defined by the IRRI Standard Evaluation System (IRRI, 2002) and CPVO's technical protocol for rice (Community Plant Variety Office, 2012). The phyllochron (Phyl, °C day$^{-1}$) was calculated as the average thermal time for complete leaf emission. Leaf emission rate was monitored on the main culm by weekly visual assessment in one replication on five plants, considering a leaf formed when the collar was visible according to Counce *et al.* (2000). Days to heading (Heading, days) was recorded as the number of days from emergence to heading (i.e., when 50% of the shoots showed emerging panicle). Days to maturity (Maturity, days) was recorded as the number of days from heading to maturity, which was visually estimated based on the toning of the hulls as a morphological marker, i.e. when all the spikelets turned brown at the end of the dry-down processes (around 22% of relative humidity) (Counce et al., 2000). Phenology was also expressed as thermal time, i.e. by estimating the growing degree days ($GDD$, °C day$^{-1}$) from emergence to reach flowering (GDDflo, °C day$^{-1}$) and from flowering to reach maturity (GDDmat, °C day$^{-1}$). $GDD$ were calculated from the daily average air temperature ($T$, °C), according to Yan and Hunt (1999):

$$GDD = r\,(T_{opt} - T_{min})$$

with:

$$r = \left(\frac{T_{max} - T}{T_{max} - T_{opt}}\right)\left(\frac{T}{T_{opt}}\right)^{\frac{T_{opt}}{T_{max} - T_{opt}}}$$

where $T_{min}$, $T_{opt}$, and $T_{max}$ are the minimum, optimum, and maximum cardinal temperatures for rice development (°C). We used the cardinal temperatures for rice (sub-species *japonica*) reported by Sanchéz et al. (2014), i.e. 10.5 °C, 29.7 °C, 42.5 °C for Phyl; 13.5 °C, 28 °C, 36 °C for GDDflo; 20.7 °C, 24.2 °C, 31.3 °C for GDDmat. The culm length (CulmLen, cm) was measured at the end of the milk stage from the soil surface to the base of the panicle. The biomass of a single complete shoot at flowering (culm + panicle + leaves, ShootDM, g) was determined by the destructive sampling of five plants without roots, which were weighted after oven drying until constant weight (~72 hours). The number of panicles per m$^2$ (Density, panicles m$^{-2}$) was assessed one week before harvest on two samples of 0.5 m$^2$ area per plot. The final leaf number (FinLeafNum, unitless) refers to the total number of leaves produced on the main culm at full flowering, counted on the same plants monitored for calculating the Phyllochron. Leaf area index (LAI, m$^2$ m$^{-2}$) was estimated at full flowering with the PocketLAI app (Confalonieri et al., 2014), as the average of ten measurements taken randomly within the plot at about 20 cm above the ground. Panicle length (PaniLen, cm) was measured on twenty plants per plot from the panicle node (base) to its tip. Panicle dry mass (PaniDM, g panicle$^{-1}$) was determined on ten panicles, which were harvested at maturity by cutting at the panicle node, and oven-dried until constant weight. The number of secondary branches per panicle (PaniBranches, unitless) was counted on the same five-panicles sample. The dry mass of a spikelet (GrainWeight, mg) was determined on four replicates of 200 fully developed spikelets measured at ~14% of relative humidity, and then the values were adjusted to dry weight according to the method provided by the International Rules for Seed Testing (International Seed Testing Association, 2019). The total number of spikelets per panicle (GrainNum, unitless) was assessed on five randomly chosen panicles by counting the number of spikelets (both filled and unfilled). Panicle sterility (Sterility, %) was measured as the average ratio between the number of filled spikelets and GrainNum. Filled spikelets were separated from unfilled ones using the airflow of a Chinese rice huller brand "Hercules" and manually counted on the same five panicles used for PaniDM and PaniBranches determination. Finally, grain yield (Yield, t ha$^{-1}$) was derived by converting the plot yield in t ha$^{-1}$ assuming a plot area of 14.4 m$^2$ and accounting for grain moisture, determined with a thermogravimetric scale, model "Sartorius MA 150". Plots were harvested with a plot combine, brand "Iseki" (Japan) model "HFC325".

## 2.4 Data analysis

The coefficients of variation (CVs) were calculated as the ratio between the pseudosigma ($\tilde{\sigma}$, a robust measure of dispersion calculated as IQR / 1.35, where IQR = inter-quartile range) (Hoaglin et al., 1983) and the median ($\tilde{x}$). We classified the CVs into low [$CV \leq (\tilde{x} - \tilde{\sigma})$], medium [$(\tilde{x} - \tilde{\sigma}) < CV \leq (\tilde{x} + \tilde{\sigma})$], high [$(\tilde{x} + \tilde{\sigma}) < CV \leq (\tilde{x} + 2\tilde{\sigma})$], and very high [$CV > (\tilde{x} + 2\tilde{\sigma})$], according to the median and pseudosigma of their distribution, as proposed by Costa *et al.* (2002), to support the analysis of the results.

For Principal Component Analysis, we adopted the methodology proposed by Husson *et al.* (2010) which integrates "illustrative elements", i.e. supplementary variables and individuals which are introduced after the computations of PCs. We used the days from emergence to heading (Heading) and the days from heading to maturity (Maturity) as supplementary quantitative variables, to avoid redundancy with GDDflo and GDDmat. The European market classification of grain shape was introduced as a supplementary qualitative variable (i.e. "Grain shape") to provide insights in the interpretation of the results, given its primary role in influencing rice selling prices. The corresponding classes were long B, round, medium, and long A (European Parliament, 2013). Long A was further divided in two categories: those intended for the parboiling process (long A-PB) and those mainly sold in the domestic market (long A-DM, i.e. traditional cultivars suitable for the preparation of typical Italian dishes like *risotto*). This distinction was adopted because of the vast phenotypic differences present between these two groups (Mongiano et al., 2018). A Hierarchical Clustering was then performed on Principal Components, retaining only the first three components to minimise the noise in data. Three main clusters were selected, maximising the relative loss of inertia (Husson et al., 2010).

We performed all the statistical analyses with the *R* statistical software (R Core Team, 2017), using the *FactoMineR* package for Principal Component Analysis and Hierarchical Clustering on Principal Components (Lê et al., 2008). Charts were created with the *ggplot2* (Wickham, 2016) and corrplot (Wei and Simko, 2017) R packages.

## 3 Results

### 3.1 Traits variability

The summary statistics of the distributions of the fourteen yield-related traits and Yield, regarding the 40 Italian rice cultivars are reported in Table 2. A boxplot representation of these data, divided by the two experimental seasons, is available in Supplementary Figure S3.

Table 2 – Summary statistics calculated for each of the considered traits. Min. = minimum value; 1Q = first quartile; 3Q = third quartile; max. = maximum value; range = distance between minimum and maximum value; SD = standard deviation; CV = coefficient of variation; Rating = classification of CVs according to the method proposed by Costa et al. (2002). List of abbreviations and measure units are reported in Table 1.

| Trait | Unit | Min. | 1Q | 3Q | Max. | Range | Mean | SD | Median | CV | Rating |
|---|---|---|---|---|---|---|---|---|---|---|---|
| **Phyl** | °C day$^{-1}$ | 54.4 | 59.4 | 78.0 | 89.9 | 35.5 | 68.4 | 10.6 | 66.3 | 20.8% | medium |
| **Heading** | days | 80.0 | 93.0 | 102.0 | 118.0 | 38.0 | 97.5 | 6.7 | 98.0 | 6.8% | low |
| **GDDflo** | °C day$^{-1}$ | 776.0 | 971.2 | 1098.5 | 1324.1 | 548.2 | 1039.1 | 95.5 | 1054.5 | 8.9% | medium |
| **Maturity** | days | 37.0 | 46.0 | 58.8 | 74.0 | 37.0 | 52.1 | 8.9 | 50.5 | 18.7% | medium |
| **GDDmat** | °C day$^{-1}$ | 329.1 | 377.7 | 412.2 | 438.8 | 109.7 | 392.4 | 26.7 | 390.3 | 6.6% | low |
| **CulmLen** | cm | 48.0 | 67.0 | 91.8 | 134.0 | 86.0 | 79.5 | 18.3 | 75.0 | 24.4% | medium |
| **ShootDM** | g | 1.2 | 2.9 | 3.9 | 6.0 | 4.8 | 3.5 | 0.9 | 3.4 | 22.6% | medium |
| **Density** | shoots m$^{-2}$ | 174.0 | 309.0 | 419.0 | 784.0 | 610.0 | 376.7 | 105.8 | 361.0 | 22.6% | medium |
| **FinLeafNum** | unitless | 10.0 | 12.0 | 13.0 | 13.0 | 3.0 | 12.3 | 0.8 | 12.5 | 5.9% | low |
| **LAI** | unitless | 3.1 | 4.1 | 4.8 | 6.4 | 3.3 | 4.5 | 0.6 | 4.4 | 10.5% | medium |
| **PaniLen** | cm | 12.0 | 17.0 | 22.0 | 27.0 | 15.0 | 19.6 | 3.2 | 20.0 | 18.5% | medium |
| **PaniDM** | g | 0.9 | 3.2 | 4.3 | 6.5 | 5.6 | 3.8 | 1.0 | 3.8 | 22.1% | medium |
| **PaniBranches** | unitless | 6.0 | 10.0 | 12.0 | 15.0 | 9.0 | 10.8 | 1.8 | 11.0 | 13.5% | medium |
| **GrainWeight** | mg | 11.0 | 24.2 | 32.2 | 46.7 | 35.6 | 28.2 | 6.0 | 27.0 | 22.0% | medium |
| **GrainNum** | unitless | 40.0 | 114.8 | 159.2 | 240.6 | 200.6 | 138.4 | 34.0 | 137.1 | 24.0% | medium |
| **Sterility** | % | 2.8 | 6.4 | 11.7 | 21.8 | 19.0 | 9.2 | 3.9 | 8.7 | 45.4% | very high |
| **Yield** | t ha$^{-1}$ | 2.2 | 5.9 | 7.3 | 9.0 | 6.8 | 6.5 | 1.2 | 6.7 | 15.5% | medium |

The thermal requirements to reach the main phenological stages (Heading - GDDflo, and Maturity - GDDmat) showed medium-low relative variation (CVs of 8.9% and 6.6%, respectively). We observed more significant variations in the duration of Heading (range 38 days) and Maturity (range 37 days) likely as the result of the interseasonal weather variability, especially in the initial stages of crop establishment and during ripening (see Methods section 2.2 and Supplementary Figure S1). The traits associated with vegetative development (Phyl, LAI, CulmLen, ShootDM) obtained medium to high CVs (20.8%, 10.5%, 24.4%, 22.6%, respectively), except FinLeafNum that had the lowest CV (5.9%). The CVs of the other traits associated with reproductive development were classified as medium to very high: the most variable trait was Sterility (CV = 45.4%) followed by Density (22.6%). We recorded a medium CV for LAI (10.5%), whose values ranged from 3.13 to 6.43. The CVs of grain features, i.e. GrainNum (CV = 24.0%), PaniDM (CV = 22.1%), and GrainWeight (CV = 22.0%) were also classified as "medium", mostly because of the large heterogeneity within the sample regarding grain shape.

The measured yield varied between 2.19 t ha$^{-1}$ (Megumi) to 8.95 t ha$^{-1}$ (Italpatna, Supplementary Figure S4); the corresponding CV was medium (15.5%), with a global average of 6.47 t ha$^{-1}$ which is consistent with the national statistics (Ente Nazionale Risi, 2018). The seasonal variability in phenotypic expression during the two years of trial was limited, with very similar distributions for most of the considered traits (Supplementary Figure S3). The largest differences were observed in traits related to phenology, especially Phyl (18 °C day$^{-1}$ median difference) and GDDflo (70 °C day$^{-1}$ median difference). Although the duration of the reproductive phase expressed as the number of days was generally longer in 2017, the corresponding cumulated thermal time remained stable, despite showing larger variability (IQR = 59 °C day$^{-1}$ in 2017 compared to 15 °C day$^{-1}$ in 2016). The remaining traits were stable in the two seasons, even those which can be tipically more affected by environmental conditions such as Sterility or Yield.

The variability of the traits Heading, Maturity, CulmLen, PaniLen, and GrainWeight within our panel of varieties was compared with the available information on the Italian rice germplasm with Brown-Forsythe test for equality of group variances (Supplementary Table S2), using the data from a previous characterisation (Mongiano et al., 2018, Supplementary Figure S2). Significant differences (p = 0.0085) were found only for Maturity, for which variance was larger in our panel than in the population, with a comparable range of variation, and Heading (p = 0.0455) for which the differences were significant at the limit of the α level considered. The variability of the other traits was not significantly different between samples, with slight differences in minimum and maximum values.

### 3.2 Traits relationships

We investigated the relationships among traits by computing the pairwise Spearman's rank correlation coefficients (ρ) and then testing their significance at α = 0.05 (Fieller et al., 1957). The correlation matrix, reporting only significant values, is provided in Figure 1.

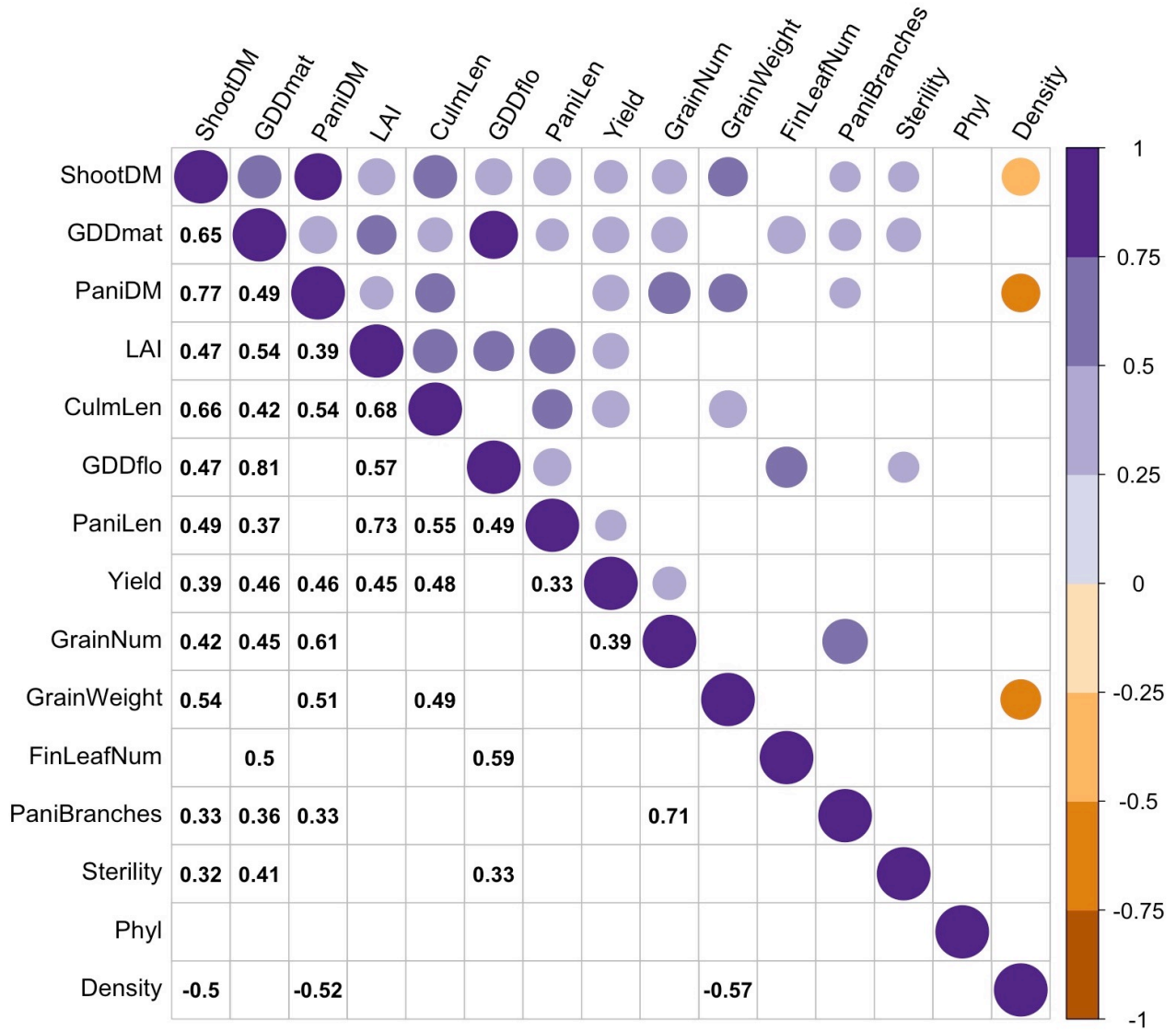

Figure 1 - Pairwise Spearman's rank correlation coefficients calculated for the considered traits. Upper diagonal: graphical representation, with the colors and size of the circles referred to the direction and intensity of the correlations. Lower diagonal: numerical coefficients. Only significant values are reported. The acronyms of the crop traits are explained in Table 1.

All the significant relationships were positive except for Density, that negatively correlated with GrainWeight (ρ = -0.57), PaniDM (ρ = -0.52) and ShootDM (ρ = -0.5). Yield was positively associated with CulmLen (ρ = 0.48), GDDmat (ρ = 0.46), PaniDM (ρ = 0.46), LAI (ρ = 0.45), GrainNum (ρ = 0.39), and ShootDM (ρ = 0.39). The strongest relationship was found between GDDflo and GDDmat (ρ = 0.81), suggesting that late flowering cultivars generally presented a long reproductive phase. GDDmat was strongly related with most traits (except GrainWeight, Phyllochron, and Density), and especially with ShootDM (ρ = 0.66), which in turn was positively associated with CulmLen ($\rho$ = 0.66) and PaniDM ($\rho$ = 0.77). As expected, PaniDM was correlated with other panicle traits, i.e. GrainNum ($\rho$ = 0.61), GrainWeight ($\rho$ = 0.51), PaniBranches ($\rho$ = 0.33), while ShootDM positively influenced most panicle traits, i.e. GrainWeight ($\rho$ = 0.54), PaniLen ($\rho$ = 0.49), GrainNum ($\rho$ = 0.42), and PaniBranches ($\rho$ = 0.33). Also, Sterility was positively associated with ShootDM ($\rho$ = 0.32), even if the strongest positive relation emerged with GDDmat ($\rho$ = 0.41). We acknowledge the limitations of the chosen statistical test, which may be prone to the residual confounding effect from the duration of vegetative and reproductive phases; considering this, the residuals of the linear correlation between growth duration, expressed in thermal time (GDDtot = GDDflo + GDDmat), and Yield were compared with the measured traits (Supplementary Figure S5). We did not find any strong pattern in the analysis of the residuals except for a slightly increasing trend and higher $r^2$ in FinalLeafNum ($r^2$ = 0.28), PaniLen ($r^2$ = 0.11), and LAI ($r^2$ = 0.10). Except for PaniLen, for which the data distribution was homogeneous along its range of variation, these trends could have been determined by the lower density of values in the bottom end of the distributions of FinalLeafNum (only three observations under 12) and LAI (only three observations over 5). We further explored the associations among traits with Principal Component Analysis (PCA), to assess the strength and direction of correlations between the original traits and the extracted Principal Components (PCs). The first three components, explaining 64.2% of the total variance, were retained for analysis (Supplementary Figure S6). Cultivar Megumi emerged as an outlier due to extreme values for most traits (Yield, GrainWeight, GrainNum, CulmLen, PaniLen, GDDflo, GDDmat) and was considered as a supplementary individual. PCA biplot of variables is reported in Figure 2, while the correlation coefficients calculated between traits and the extracted PCs are reported in Supplementary Table S3.

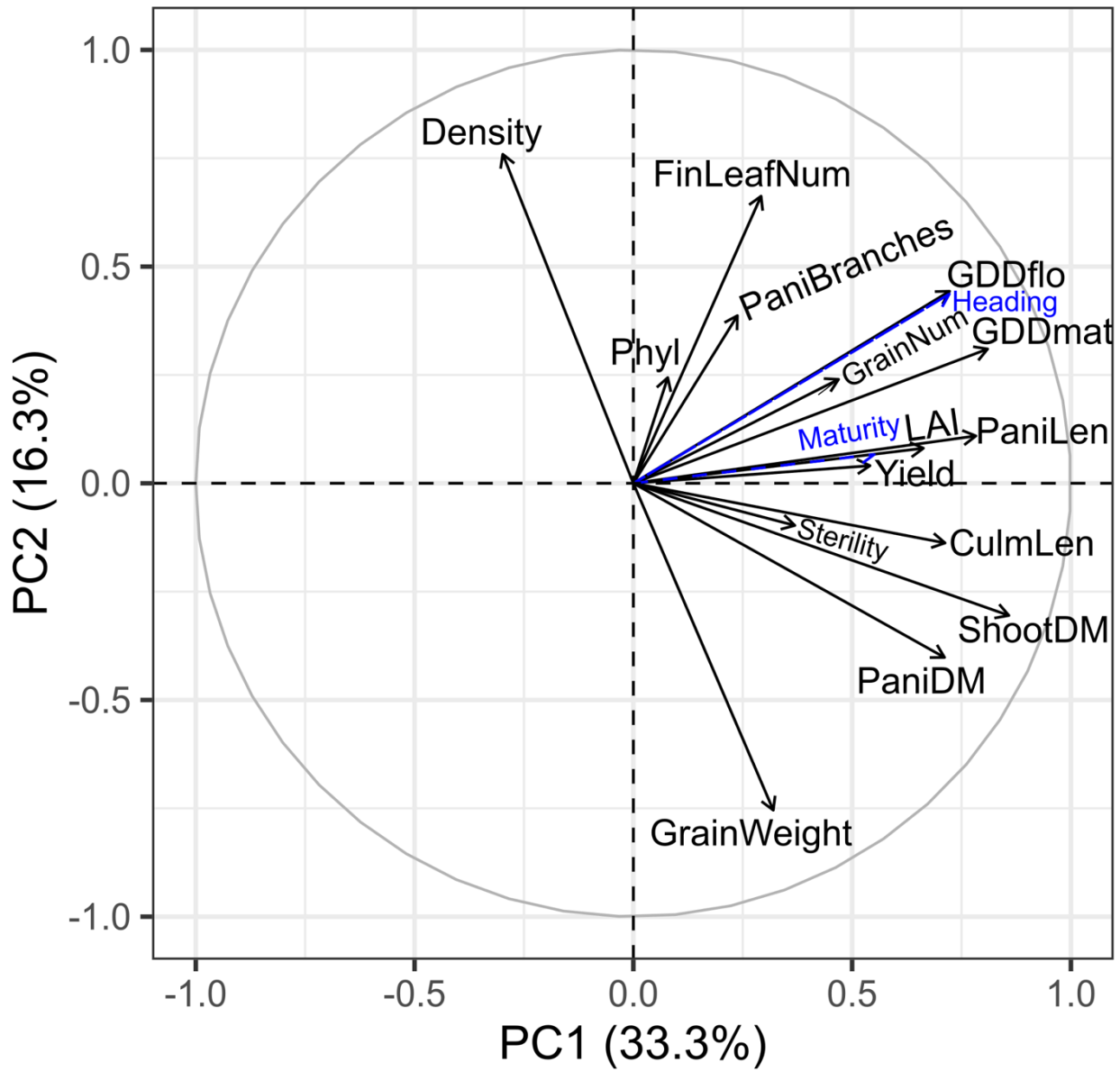

Figure 2 - PCA biplot of variables. List of abbreviations is reported in Table 1. Supplementary variables (i.e. illustrative elements, not taking parts in the computation of principal components space) are colour-coded in blue.

PC1 synthesised the direct relations between most traits and Yield. We found the strongest correlations with PC1 for ShootDM (0.86), GDDmat (0.81), LAI (0.78), GDDflo (0.72), CulmLen (0.71), PaniDM (0.71), PaniLen (0.66), and Yield (0.54). PC2 was positively correlated with Density (0.76), FinalLeafNum (0.66), GDDflo (0.44), and PaniBranches (0.39), while we found negative correlations with PaniDM (-0.40) and GrainWeight (-0.75). Cultivars at positive coordinates on PC2 had a longer duration of the vegetative phase, with a higher number of leaves and increased tillering, but lighter seeds and panicles. PC3 positively correlated with Phyl (0.46), GrainWeight (0.33), and FinLeafNum (0.35), and negatively with GrainNum (-0.77), PaniBranches (-0.72), and PaniDM (-0.40). These relations suggest that a higher final number of leaves, larger phyllochron, increased panicle length and seeds weight led to fewer secondary rachis-branches and fewer seeds per panicle, associated with lower panicle biomass (and *vice versa*). The differences between the mean coordinates of cultivars grouped by grain shape and the overall mean were tested for significance (*t-tests*, Supplementary Table S4). The results indicated that long A-DM group had significantly different coordinates on PC1 (positive), PC2 (negative), and PC3 (positive), suggesting that these cultivars showed considerable biomass accumulation during both vegetative and reproductive phases, while producing few tillers, less branched panicles and fewer but heavier grains, with an extended phyllochron and higher final number of leaves. Long A-DM opposed to long B grain cultivars on PC2, the latter presenting significantly lighter grains with a higher number of grains carried by longer panicle with many ramifications; furthermore, they produced more leaves and tillers.

### 3.3 Cluster analysis

Cluster analysis was conducted to highlight grouping patterns among the tested cultivars and to provide a phenotypic classification (Figure 3).

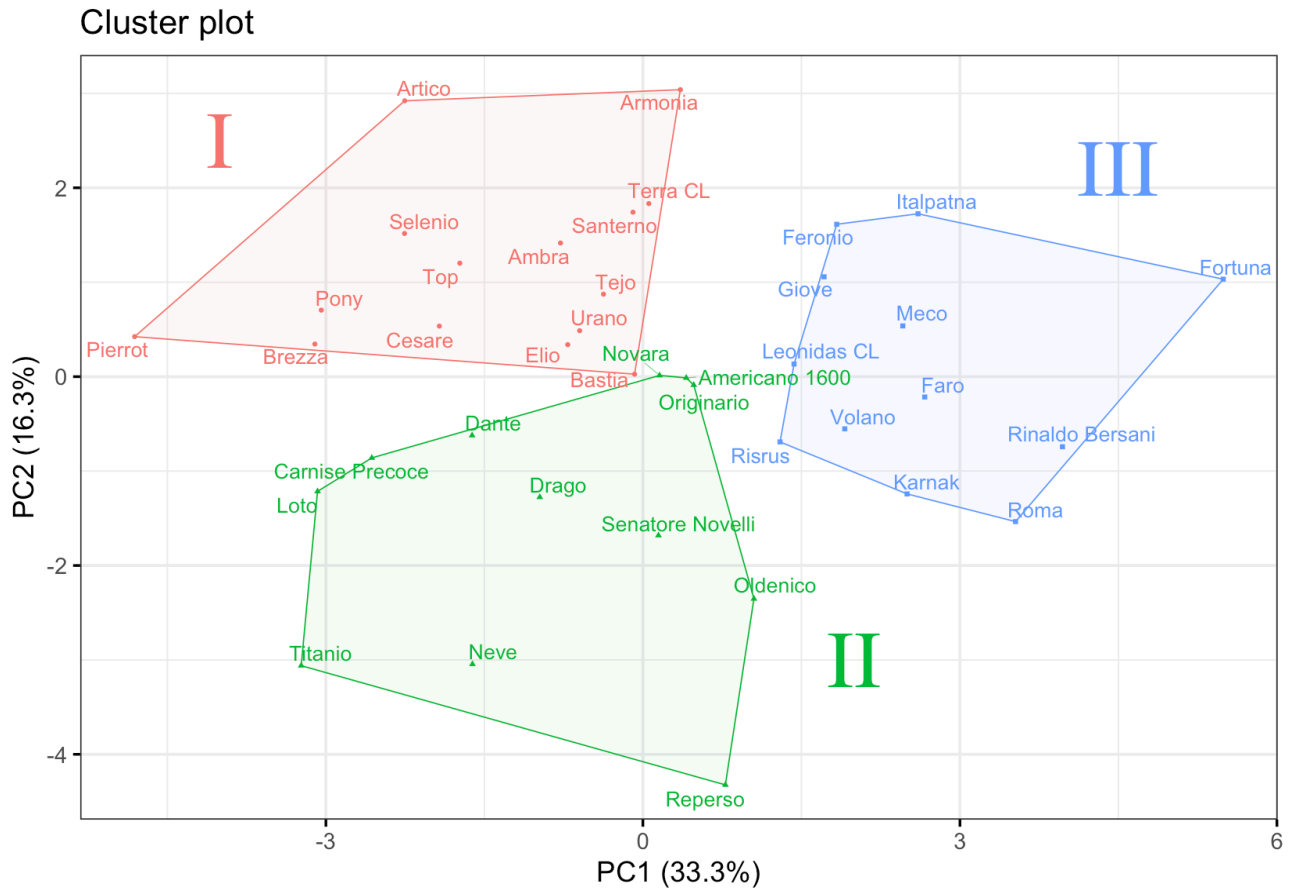

Figure 3 - PCA biplot of individuals with an indication of the three extracted clusters resulting from Hierarchical Clustering on Principal Components (HCPC).

We assessed the between-cluster variance over the total variance explained by each trait by calculating $\eta^2$ (Supplementary Table S5). The variables explaining the largest amount of between-cluster variance were GDDmat ($\eta^2 = 0.54$) and GDDflo ($\eta^2 = 0.54$), ShootDM ($\eta^2 = 0.47$), and PaniDM ($\eta^2 = 0.41$). The value of $\eta^2$ calculated for Yield was low ($\eta^2 = 0.17$), suggesting that this trait was not highly discriminant between clusters, which evenly grouped cultivars from the whole range of variation of yield.

Cluster I comprised all the long B cultivars, except for Giove, including three short-height long A for parboiling (Cesare, Pony and Tejo) and several round grain cultivars (Ambra, Bastia, Elio, Pierrot, Selenio, Terra CL, Top). This cluster was characterised by high-tillering, short-height cultivars that presented mean values of GrainWeight, PaniLen, PaniDM, and ShootsDM lower than average. The mean Yield in this cluster was slightly lower than the overall mean (6.16 t ha$^{-1}$ compared to 6.57 t ha$^{-1}$), despite the presence of modern high-yielding cultivars, like Selenio and

Terra CL. Terra CL (medium-height, late, round grain cultivar) placed far from the barycentre of the cluster, while cultivar Top (short-height, early, round grain) was the closest.

In Cluster II, most of the individuals were early long A cultivars except for three historic accessions released at the beginning of 20$^{th}$ century, i.e. Novara (medium), Americano 1600, and Originario (round). Cultivars in Cluster II presented above average GrainWeight but reduced PaniBranches, Density, Sterility, GDDmat, GDDflo and FinLeafNum. Cultivar 'Senatore Novelli' (traditional low-tillering genotype with long culm and intermediate time of maturity) was one of the closest to the barycentre of the cluster; on the contrary, cv. Titanio (long A for parboiling grain, very early and with limited tillering attitude) was at the farthest distance from the barycentre. This cluster comprised cultivars with short cycle duration, fewer leaves and tillers, and low sterility and panicles with fewer secondary rachis-branches. The lowest yielding cultivars from this cluster (Neve 5.58 t ha$^{-1}$, Carnise precoce 5.98 t ha$^{-1}$, Titanio 5.26 t ha$^{-1}$) were characterised by minimum thermal requirements for the vegetative and reproductive stage. On the contrary, cultivars Americano 1600 and Originario (average yield above 7 t ha$^{-1}$) had an extended crop cycle. Cluster III grouped rice cultivars with many traits above the overall mean, i.e., GDDmat, GDDflo, ShootDM, LAI, PaniDM and PaniLen, CulmLen, FinLeafNum, and GrainWeight. An above-average Sterility (11%) also characterised Cluster III. Half of the long A (DM) cultivars were included, except for one round grain and one long B cultivar. The remaining were long A-PB. Genotypes in this cluster were characterised by an extended crop cycle associated with a remarkable biomass accumulation (longer culms, more leaves, bigger panicles and seeds) both in the vegetative and reproductive phase. These cultivars were all high-yielding with the exception of Fortuna (5.41 t ha$^{-1}$), which is an accession probably imported from Louisiana during the first years of 1900 and later adapted to Italian environments (Adair et al., 1973) characterised by the largest thermal requirement for vegetative growth and a remarkable biomass accumulation before heading. The composition of cultivars in this cluster further confirmed that an increased source and sink capacity is associated with larger thermal requirements, often leading to higher yield.

# 4 Discussion

## 4.1 Portraying the phenotype space of the panel

We selected a panel of 40 cultivars from 351 Italian genotypes which were characterised in a previous study for phenology (days to heading and days to maturity), culm and panicle length, thousand seeds weight, caryopsis width and length (Mongiano et al., 2018). The variability found in the 40 Italian rice cultivars was considerable for all the analysed traits and consistent with published data (Faivre-Rampant et al., 2010; Katsura et al., 2007; Samonte et al., 2001; 1998; Volante et al., 2017). This result suggests that the cultivar selection method employed (Kennard-Stone algorithm) was effective in extracting a representative sample from a wide population of 351 Italian cultivars, with large variability in the crop traits. Unfortunately, we do not have any measured data to verify that our sample was representative of the variability of the new traits characterized in the present study. It is, however, realistic to hypothesize that a considerable part of variability was captured also for these traits given the multiple correlations emerged both in the pairwise correlation analysis and in the PCA, and the consistency with the ranges from literature.

In our experiment, the weather variability in the two seasons did not have a significant impact in modulating the phenotypic expression of the cultivars. Minor differences were observed for phenological traits, especially Phyl, although in line with was reported in the literature and likely attributable to seasonal differences in temperatures during early vegetative development (Wilhelm and McMaster, 1995).

## 4.2 Investigating the relationships among crop traits

The traits associations were evaluated a) via pair-wise Spearman's rank correlation coefficients, b) with Principal Components Analysis, and c) by observing patterns highlighted by Cluster

Analysis. As expected, panicle weight (i.e. the sink size) was significantly associated with yield. The former depends upon the interaction of grain weight, number of grains per panicle, length of the primary axis of the panicle and the number of secondary rachis-branches, that are often defined as "panicle components" (Hittalmani et al., 2003). Highest-yielding cultivars presented different combinations of these traits, that led to heavier panicles (Dingkuhn et al., 2015). Other than being related to the characteristics of grains, higher yields were also associated with extended thermal requirements to reach flowering (GDDflo) and maturity (GDDmat) stages, likely because they are associated with larger accumulation of solar radiation (Katsura et al., 2007). Increased aboveground biomass (ShootDM and PaniDM) was indeed associated with extended vegetative and reproductive phases, which were strongly linked and positively correlated to many other traits impacting yield, i.e. CulmLen, PaniLen, FinLeafNum, and LAI, as pleiotropic effects are known to exist amongst the genes controlling these traits (Xue et al., 2008).

Another primary outcome of our study is that the rice cultivars maintained a constant balance between the biomass accumulated during the vegetative and ripening stage, confirmed by the significant association between ShootDM and PaniDM. This finding is in agreement with previous studies that highlighted the importance of non-structural carbohydrates accumulation before heading, which contributes to a large portion of grain carbohydrates and serves as a buffer during periods of sub-optimal radiation levels (Katsura et al., 2007; Samonte et al., 2001; Stella et al., 2016).

### 4.3 Yield strategies in the Italian rice cultivars

Compensatory mechanisms are known to modulate the sink-source balance (Kumar et al., 2016). Genotypes with short culms accumulated less biomass during the vegetative stage, had smaller leaf area, lighter panicle, and had a short cycle duration, resulting in low grain yield. Examples of these genotypes are Megumi, Pierrot, and Pony, all short-culm, high-tillering and very early genotypes that produced an average grain yield lower than 5 t ha$^{-1}$. On the contrary, most of the highest-yielding genotypes accumulated a large amount of biomass during the whole crop cycle, due to either long culms and a high number of leaves (Novara, Americano 1600), or to shorter culms but of increased dry mass (Meco), or to an increased number of productive tillers (Italpatna, Terra CL), fine-tuned according to their panicle component traits, like GrainNum and GrainWeight (Katsura et al., 2007; Peng et al., 2008). All these features were supported by an extended crop cycle, while earliness was often associated with lower yields (Vergara et al., 1964). One exception was the cultivar Novara that, despite being one of the genotypes with the shortest duration of the crop cycle led to an average yield of 7.51 t ha$^{-1}$. A possible explanation relies in the favourable ratio between the duration of the ripening stage over the whole crop cycle (28.7%), associated with high biomass accumulation during the vegetative stage, which is typical of high-yielding genotypes (Peng et al., 2000). When earliness was associated with an increased number of tillers, it showed to negatively affect yield, probably due to the increased abortion phase (usually occurring after panicle initiation) determined by the limited biomass accumulation (Kumar et al., 2016). Early flowering genotypes presented many unproductive tillers, counterbalanced by reduced seed and panicle size as the partitioning of the assimilates to tillers impacted biomass accumulation in individual shoots (Mousanejad et al., 2010). In our experiment, when this trait was highly expressed (Pony - Long A, Brezza - Long B, Pierrot - Round), rice yield was lower than average (4.99, 5.18, and 3.68 t ha$^1$ respectively). There are, however, cultivars like Ambra and Selenio (round) that produced average grain yield higher than 7 t ha$^1$, despite their high-tillering attitude. Contrarily to low-yielding cultivars with enhanced tillering ability, their primary differences were the duration of the vegetative and reproductive phases, which is known to be crucial in supporting the growth of an increased number of tillers (Dingkuhn et al., 1991). Other examples of these cultivars are Artico and Armonia (long B), or Tejo (long A). GDDmat was also strongly correlated with Sterility, which was more severe in late genotypes probably because of the lower temperatures typical of the late season, leading to not fully developed panicles. Although we cannot vprovide any quantitative explanation, this hypothesis is supported by the fact that most of the empty spikelets were at the base of panicle, since anthesis and ripening occur in a basipetal gradient (Counce et al., 2000) and spikelets in inferior positions have been reported to be more susceptible to sterility due to cold temperatures (Chen et al., 2017). However, spikelet sterility did not significantly impacted on yield in our dataset, as many of the high-yielding cultivars (Faro, Volano, Meco) also presented marked Sterility. We assume that in our experiment sterility was mainly driven by genetic factors in response to the environment like e.g. period of anthesis, time of day of anthesis or physiological tolerance like the quantity of pollen or morphology of reproductive organs (Jagadish et al., 2009), rather than the environmental component (Julia and Dingkuhn, 2013), in light of the consistent cultivar-specific response in the two seasons.

Cluster analysis extracted three groups that can be briefly summarised as 'high-tillering' (Cluster I), 'early' (Cluster II), and 'increased source-sink' (Cluster III) genotypes. The classifications found in literature for rice cultivars based on morphological descriptors mostly concern the ecotype and grain features, whereas only few studies have proposed classifications based on yield-related traits (Mathure et al., 2011; Schlosser et al., 2000; Yawen et al., 2003). Our grouping partially resemble the 'panicle number type' (Cluster I) and 'panicle weight type' (Cluster III) groups to which some authors refer to generally indicate ideotypes or breeding strategies (IRRI, 1991; Katsura et al., 2007; Peng et al., 2008; Takai et al., 2019), although the classification provided here relies on multiple traits other than panicle dry mass or density. Cultivars at the farthest location from the barycentre of their respective cluster were either low- or high-yielding cultivars. Lower yields were associated to genotypes at extreme coordinates (both positive or negative) on PC1 (Pierrot, Pony, Brezza, Loto, Titanio, Fortuna), while many of the high-yielding genotypes were at coordinates close to the origin on PC1 (Terra CL, Novara, Americano 1600, Originario). This result suggests that when a specific "breeding strategy" is pursued, a proportionate ratio between yield-related traits has to be maintained for high grain production, i.e. not uni-directionally stretched to the limits of the phenotypic range of the involved crop traits (Katsura et al., 2007). Our results also confirmed previous characterisations of Italian rice germplasm (Faivre-Rampant et al., 2010; Mongiano et al., 2019), which found out that cultivars sharing similar grain shape also displayed similar phenotypic traits. Specifically, most of the 'increased source-sink' genotypes were found to be long A, and among the "high-tillering" the 80% of cultivar belong to 'long B' or 'round' grain shape. 'Early' genotypes were instead more heterogenous as this yield strategy is shared by all grain shape-based groups.

## 4.4 Assumptions and perspectives

Three pre-mechanisation genotypes, i.e. bred before 1960 (Novara, 1930; Americano 1600, 1921; Originario, 1904), were amongst the top-yielding cultivars and no linear relationship between year of release and yield was detected in our dataset ($R^2 = 0.003706$, $p = 0.2973$), although the cultivars included in the study were not selected on the basis of their productivity. Nevertheless, this unexpected result is in agreement with studies from different geographical regions, that associated most of the recent yield increases to better cultivation techniques and increased fertiliser inputs rather than to genetic improvement (Dingkuhn et al., 2015; Peng et al., 2000). Rice growers have stopped growing these genotypes many decades ago because of their undesirable agronomic traits like high susceptibility to rice blast (*Magnaporthe grisea* - T.T. Hebert - M.E. Barr), the tendency to lodging, large amounts of unwanted residual straw and excessive duration of the growing cycle (Faivre-Rampant et al., 2010; Titone et al., 2015). Furthermore, the market demands regarding grain quality completely changed in the last century (Tamborini and Lupotto, 2006). The management practices used in our trials mitigated these unfavourable features so that their remarkable growth rate has been fully expressed. Rice blast was efficiently controlled with the use of fungicides and appropriate nitrogen inputs as excessive fertilization rates favour disease severity (Ballini et al., 2013; Bregaglio et al., 2017; Webster and Gunnell, 1992). Moreover, we did not have to account for straw management and harvest complications due to long crop cycle. Lodging was still present, even though in our trial it did not significantly impact yield. The variability in agricultural practices was completely avoided in our study since all cultivars were managed with the same practices, with a fixed sowing date. This represents a simplification because alternative management strategies markedly affect productivity. For example, late sowing could have positively affected the yields of early genotypes (Loto, Titanio), by postponing the grain filling period to more favourable conditions, as high temperatures during the ripening phase are responsible for anticipated panicle senescence, resulting in lower yields (J. Kim et al., 2011). However, the aim of the study required this simplification in order to focus on the genotypic component, so that we tried to minimise the impact of agricultural management by using standard practices adopted in the main Italian rice cultivation area.

One question that remains unanswered in the present study is whether the measured ranges of traits could be used to determine their physiological limits in the current germplasm, even if we assume that our sample was representative of the maximum variance in the measured traits of Italian rice cultivars.

## 5 Conclusion

Italian rice germplasm is a precious reservoir of genetic variability, accumulated throughout the twentieth century until today by the synergistic efforts of farmers, breeders, and scientists, further demonstrated by the high phenotypic variability found within this representative panel of cultivars.

The study analysed the ranges of variation of critical yield-related traits in Italian rice cultivars and shed lights on their associations. Results further reinforced the concept that neither single characteristics nor group of features can be indisputably associated with high yield. Instead, different yield strategies emerged from the study, resembling the efforts of plant breeders throughout the Italian rice breeding history.

The data collected in this study is ready to be used by breeders and crop modellers, who need and rely upon this type of information to design new breeding programs. What we provide is an experimental dataset of the phenotypic traits associated with the main biophysical processes implemented in crop simulation models, together with their range of variations and correlations. This research could as well foster the adoption of crop models to boost rice breeding since the definition of new ideotypes can now refer to the available phenotypic space.

# 6 Declarations

All data generated or analysed during this study are included in this published article and its supplementary information files.

## 6.1 The authors declare no conflicts of interest.

This research has been hosted in the experimental fields of CREA-DC as part of a collaboration with the University of Milan for the PhD of Gabriele Mongiano and did not receive any specific grant from funding agencies in the public, commercial, or not-for-profit sectors.

## 6.2 Acknowledgements

The authors thanks Dr Sofia Fregonara for her invaluable help in the field and laboratory activities.

## 6.3 Credit Author Statement

**Gabriele Mongiano:** conceptualization, methodology, software, validation, formal analysis, investigation, data curation, writing – original draft, review & editing, visualization. **Patrizia Titone:** writing – review & editing, supervision, project administration. **Simone Pagnoncelli:** investigation. **Davide Sacco:** investigation. **Luigi Tamborini:** supervision, project administration, funding acquisition. **Roberto Pilu**: supervision, writing – review & editing. **Simone Bregaglio:** conceptualization, writing – review & editing, methodology, validation, project administration

# Phenotypic variability in the Italian rice germplasm
## Supplementary Material

Gabriele Mongiano, Patrizia Titone, Simone Pagnoncelli, Davide Sacco,
Luigi Tamborini, Roberto Pilu, Simone Bregaglio

# Contents

# Methods

## List of field management practices

Supplementary Table S1: List of field management practices performed during the execution of field trials during both seasons.

| Date | Operation | Notes |
| --- | --- | --- |
| 2016-03-24 | Organic fertiliser application (3-1-2) | 15 g m$^{-2}$ |
| 2016-03-29 | Ploughing (~ 0.3m) | |
| 2016-04-05 | Levelling of the paddy | |
| 2016-04-13 | Harrowing | |
| 2016-04-16 | Chemical weeding (pre-sow) | cycloxydim, bensulfuron-methyl, oxadiazon |
| 2016-04-29 | Chemical weeding (pre-emergence) | clomazone, pendimethalin |
| 2016-06-03 | Mineral fertiliser application (urea, 46-0-0) | 13 g m$^{-2}$ |
| 2016-07-15 | Mineral fertiliser application (23-0-30) | 18.2 g m$^{-2}$ |
| 2016-07-26 | Fungicide application | tryciclazole, azoxystrobin |
| 2017-02-08 | Organic fertiliser application (3-1-2) | 15 g m$^{-2}$ |
| 2017-04-04 | Ploughing (~ 0.3m) | |
| 2017-04-12 | Levelling of the paddy | |
| 2017-04-22 | Chemical weeding (pre-sow) | cycloxydim, bensulfuron-methyl, oxadiazon |
| 2017-06-09 | Chemical weeding (post-emergence) | penoxsulam, quinlorac, bensulfuron-methyl, lambda-cyhalothrin |
| 2017-06-21 | Mineral fertiliser application (urea, 46-0-0) | 13 g m$^{-2}$ |
| 2017-07-03 | Chemical weeding (post-emergence) | cyhalofop-butyl |
| 2017-07-15 | Mineral fertiliser application (23-0-30) | 21 g m$^{-2}$ |
| 2017-07-26 | Fungicide application | tryciclazole, azoxystrobin |

# Minimum and maximum daily temperatures

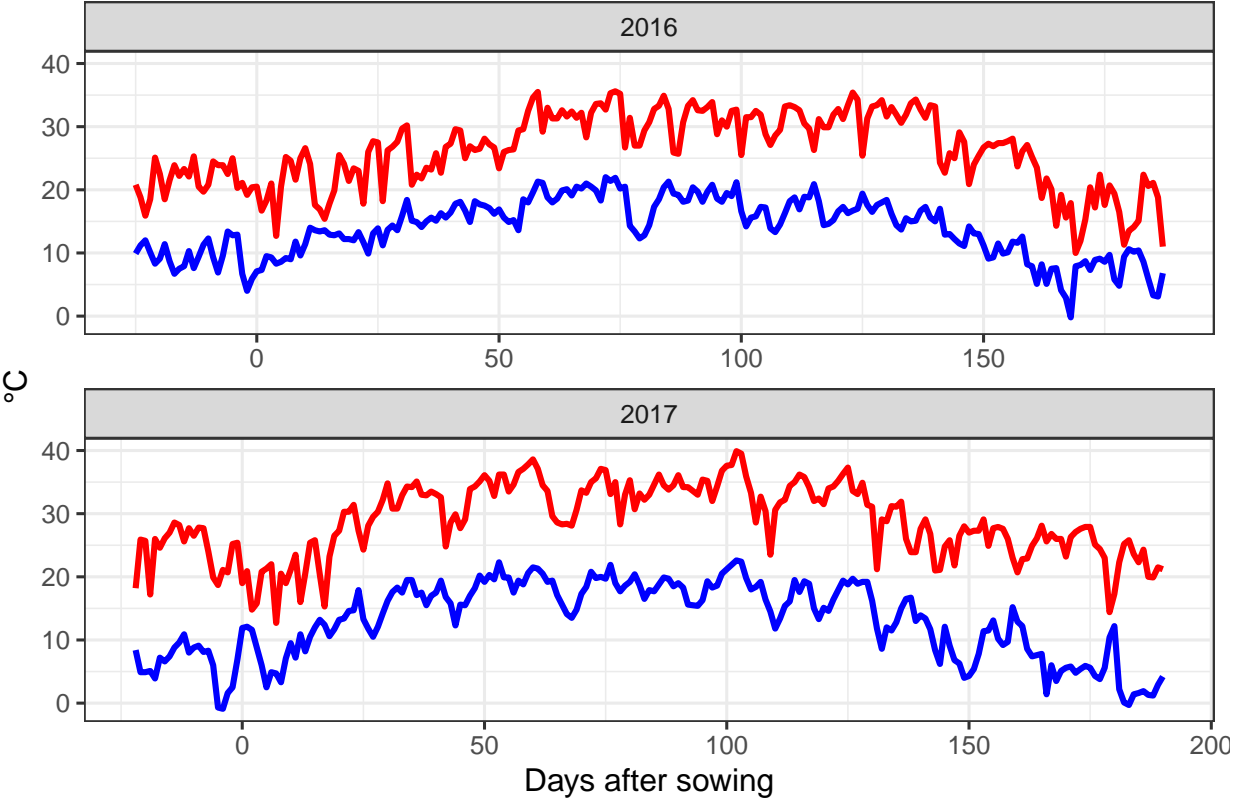

Supplementary Figure S1: Minimum (blue) and maximum (red) daily temperatures (°C) recorded in the two seasons of the trial.

# Variability of traits

## Comparison of the extracted cultivars sample with large population

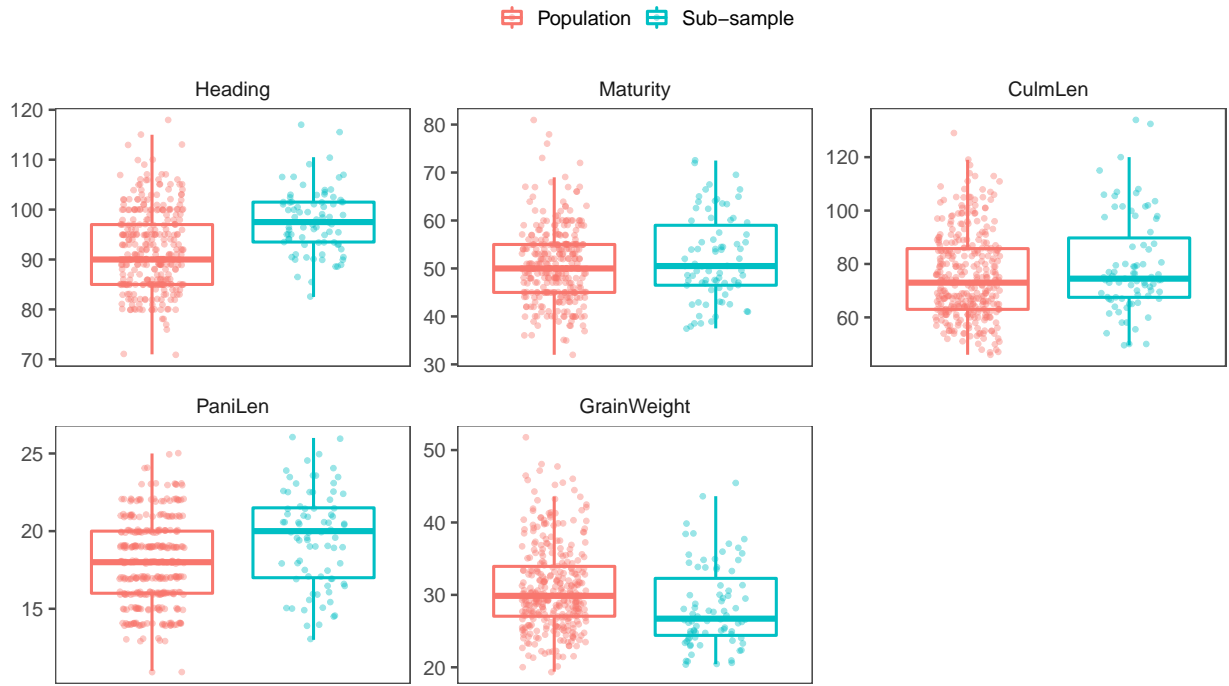

Supplementary Figure S2: Phenotypic variability within the representative panel of 40 cultivars extracted using the Kennard-Stone algorithm compared with that found in the available Italian germplasm (Mongiano et al. 2018) for traits Heading (days), Maturity (days), CulmLen (cm), PaniLen (cm), and GrainWeight (mg). Data points are horizontally jittered to minimise overlapping. Abbreviations and measure units of the studied variables are reported in Table 1

**Brown-Forsythe test for equality of variances**

Supplementary Table S2: Results of the Brown-Forsythe test for equality of variances per formed to compare the variability extracted by the sample with respect to the original population. Signif. codes: 0 '***' 0.001 '**' 0.01 '*' 0.05 '.' 0.1 ' ' 1

| Trait | Prob ($> F$) | |
|---|---|---|
| Heading | 0.0455 | |
| Maturity | 0.0085 | $**$ |
| CulmLen | 0.6526 | |
| PaniLen | 0.081 | |
| GrainWeight | 0.881 | |

# Box plots: traits distributions.

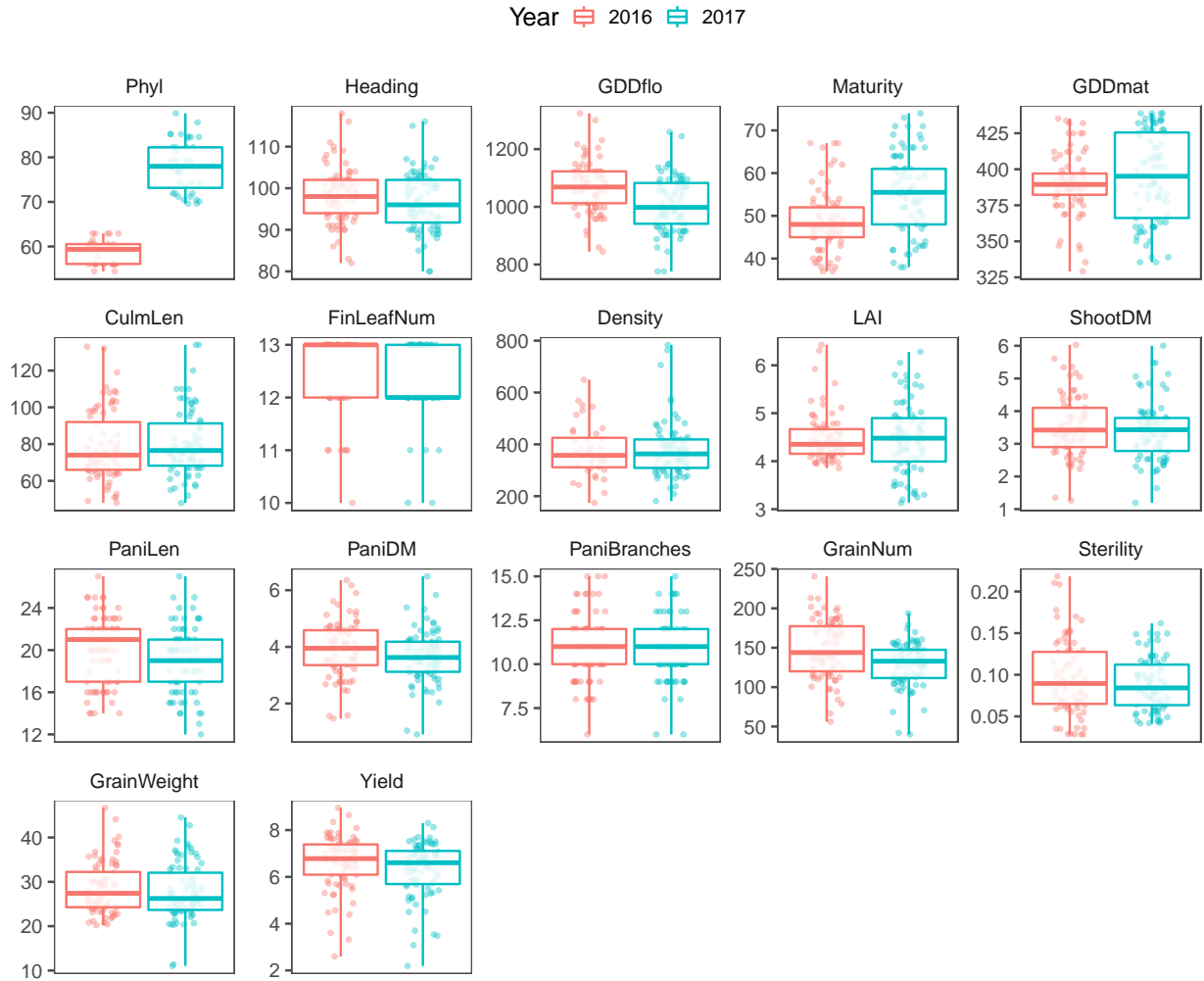

Supplementary Figure S3: Representation of traits data distribution with boxplot combined with data points, grouped by growing season. Points are jittered horizontally to minimise overlapping. Abbreviations and measure units of the studied variables are reported in Table 1

# Cultivars average yield

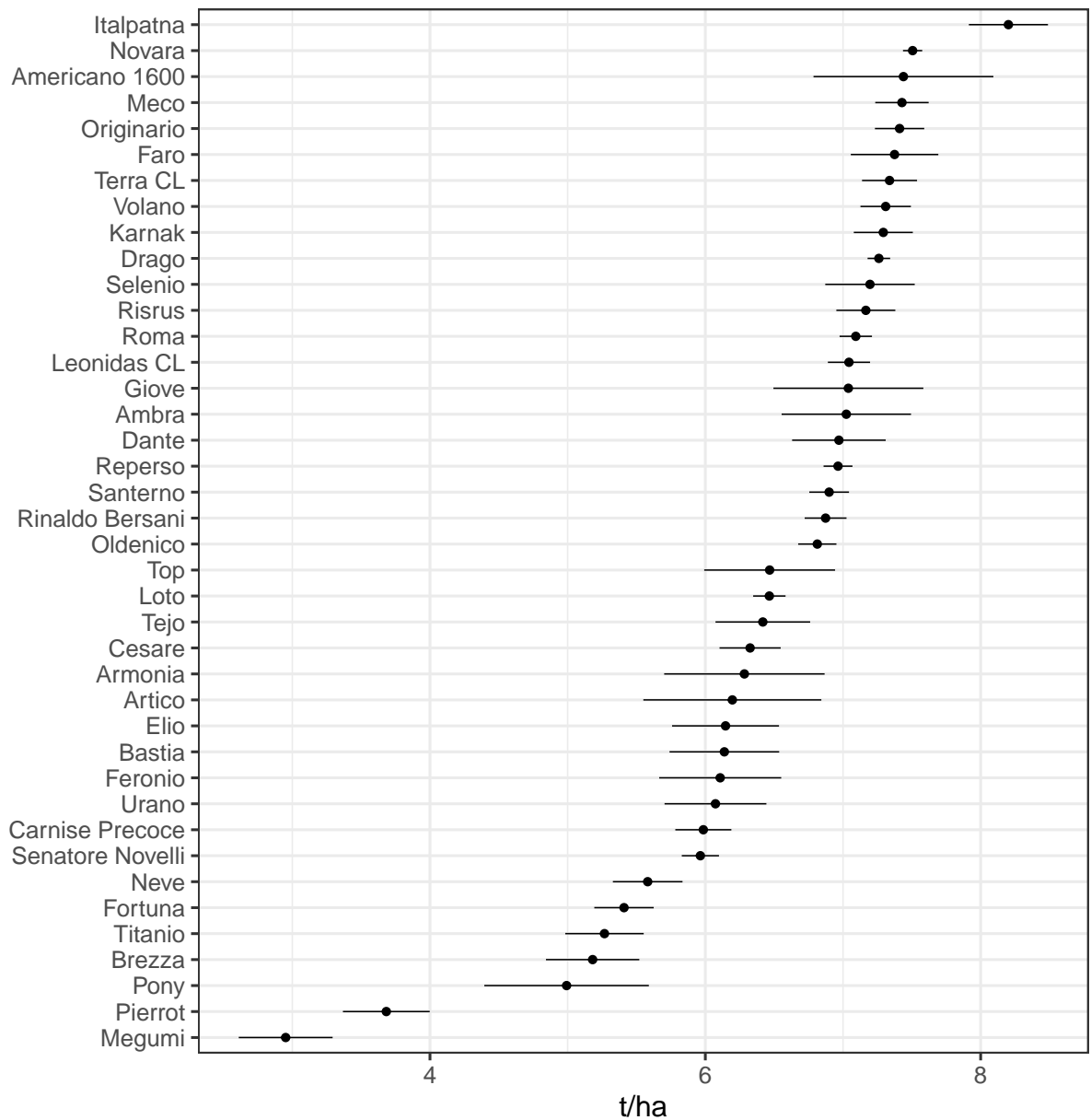

Supplementary Figure S4: Plot showing mean yield (point) and standard deviation (errorbars) measured in the two seasons experiment, expressed as $t \cdot ha^{-1}$.

## Analysis of the residuals

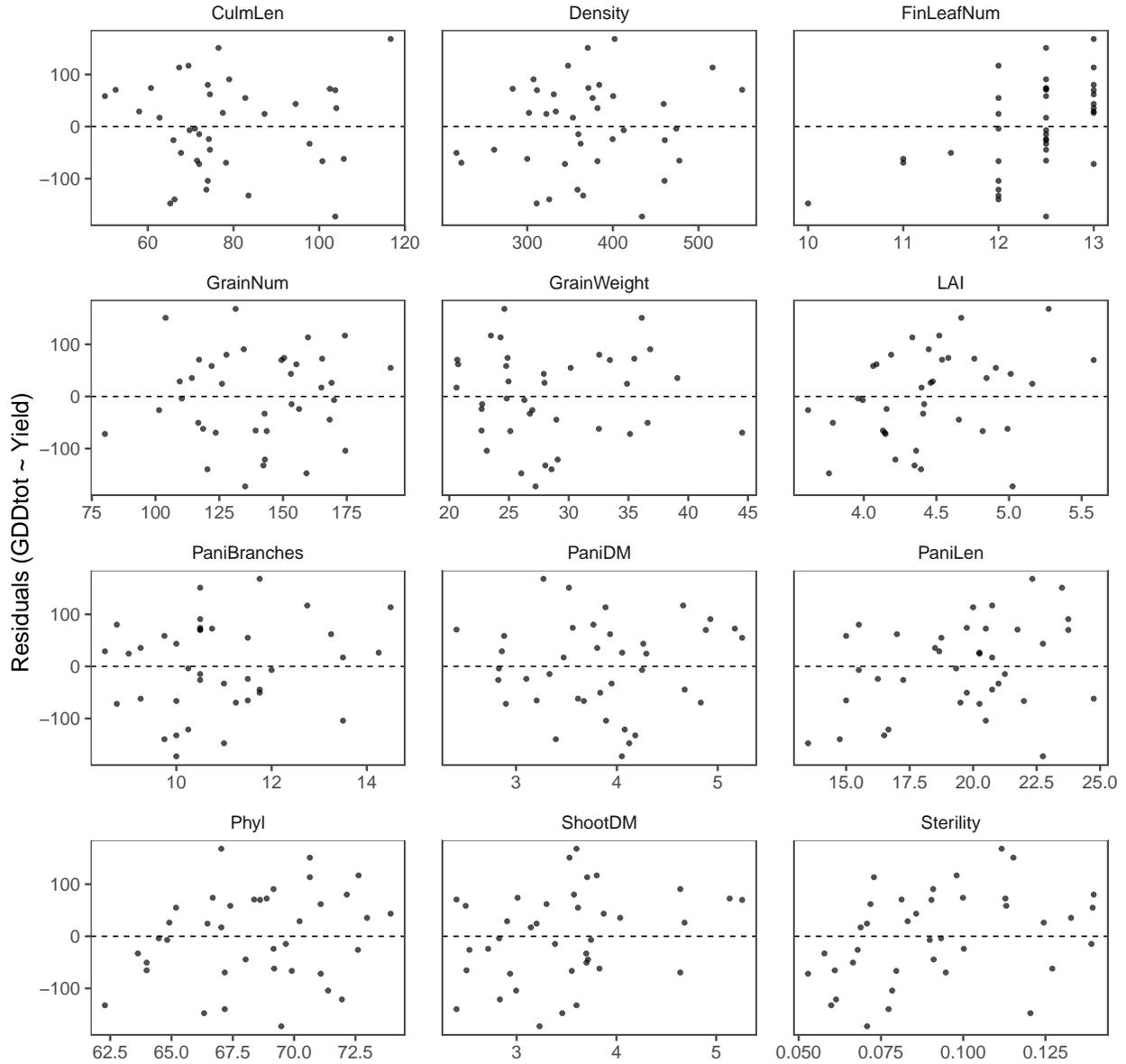

Supplementary Figure S5: Analysis of the relations between the residual of the relationship between Yield (t ha$^{-1}$) and growth duration, expressed in thermal time, (GDDtot = GDDflo + GDDmat, °C day$^{-1}$) with each of the measured traits. Abbreviations and measure units of the studied variables are reported in Table 1

# Principal Component Analysis

## Screeplot

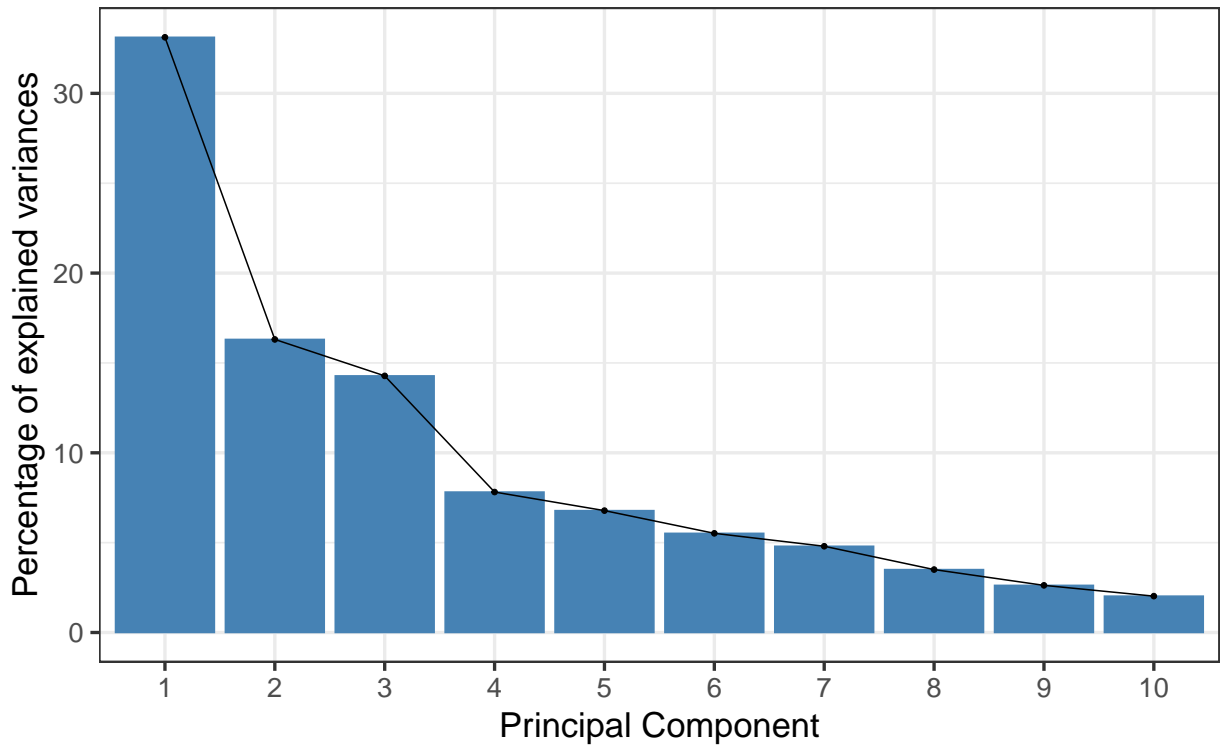

Supplementary Figure S6: Percentage of variance explained by each of the extracted Principal Components resulting from Principal Components Analysis

# Correlations between Principal Components and the original variables

Supplementary Table S3: Correlation coefficient calculated between variables and the first three Principal Components, tested for significance (Sig). Significance codes: * $p < 0.05$, ** $p < 0.01$, *** $p < 0.001$.

| Trait | PC1 | Sig | PC2 | Sig | PC3 | Sig |
|---|---|---|---|---|---|---|
| Phyl | 0.08 | | 0.24 | | 0.46 | * * |
| Heading | 0.72 | * * * | 0.44 | * * | 0.19 | |
| GDDflo | 0.72 | * * * | 0.44 | * * | 0.19 | |
| Maturity | 0.55 | * * * | 0.07 | | -0.31 | |
| GDDmat | 0.81 | * * * | 0.31 | | -0.08 | |
| CulmLen | 0.71 | * * * | -0.14 | | 0.28 | |
| ShootDM | 0.86 | * * * | -0.3 | | -0.07 | |
| Density | -0.3 | | 0.76 | * * * | 0.14 | |
| FinLeafNum | 0.29 | | 0.66 | * * * | 0.35 | * |
| LAI | 0.78 | * * * | 0.11 | | 0.32 | * |
| PaniLen | 0.66 | * * * | 0.08 | | 0.38 | * |
| PaniDM | 0.71 | * * * | -0.4 | * | -0.4 | * |
| PaniBranches | 0.24 | | 0.39 | * | -0.72 | * * * |
| GrainWeight | 0.32 | * | -0.75 | * * * | 0.33 | * |
| GrainNum | 0.47 | * * | 0.24 | | -0.77 | * * * |
| Sterility | 0.37 | * | -0.1 | | 0.03 | |
| Yield | 0.54 | * * * | 0.04 | | -0.19 | |

# Supplementary qualitative variables

Supplementary Table S4: Coordinate estimates of the categories describing the grain shape of cultivars on each Principal Components, tested for significance (Sig) with *t-test*. Significance codes: * $p < 0.05$, ** $p < 0.01$, *** $p < 0.001$.

| Category | PC1 | Sig | PC2 | Sig | PC3 | Sig |
|---|---|---|---|---|---|---|
| graintype=long A-DM | 1.688 | * * | -1.455 | * * * | 0.5827 | * |
| graintype=long A-PB | -0.4372 | | -0.1099 | | -0.3309 | |
| graintype=long B | -0.5839 | | 1.425 | * * | -0.4764 | |
| graintype=medium | 0.2374 | | -0.16 | | 0.9042 | |
| graintype=round | -0.9039 | | 0.3 | | -0.6797 | |

# Clustering

## Eta-squared

Supplementary Table S5: $\eta^2$ values calculated for the considered traits and tested for significance (Sig). Significance codes: * p < 0.05, ** p < 0.01, *** p < 0.001.

| Trait | $\eta^2$ | Sig |
|---|---|---|
| GDDflo | 0.5446 | * * * |
| GDDmat | 0.5441 | * * * |
| Heading | 0.5405 | * * * |
| ShootDM | 0.4734 | * * * |
| PaniDM | 0.4073 | * * * |
| CulmLen | 0.3867 | * * * |
| LAI | 0.3658 | * * * |
| GrainWeight | 0.3614 | * * * |
| FinLeafNum | 0.3534 | * * * |
| Density | 0.3101 | * * |
| PaniLen | 0.2472 | * * |
| Sterility | 0.2307 | * * |
| Maturity | 0.2279 | * * |
| Yield | 0.1743 | * |
| GrainNum | 0.1318 | |
| PaniBranches | 0.1229 | |
| Phyl | 0.05744 | |